# Temperature dependent Raman spectroscopy of titanium trisulfide (TiS$_3$) nanoribbons and nanosheets

*Amit S. Pawbake[1,5], Joshua O. Island[2], Eduardo Flores[3], Jose Ramon Ares[3], Carlos Sanchez[3], Isabel J. Ferrer[3], Sandesh R Jadkar[1], Herre S. J. van der Zant[2], Andres Castellanos-Gomez[4\*], Dattatray J Late[5\*]*

[1]School of Energy studies, Department of Physics, Savitribai Phule Pune University, Pune 411007, India
[2]Kavli Institute of Nanoscience, Delft University of Technology, Lorentzweg 1, 2628 CJ, Delft, The Netherlands
[3]Materials of Interest in Renewable Energies Group (MIRE Group), Dpto. de Física de Materiales, Universidad Autónoma de Madrid, UAM, 28049 Madrid, Spain
[4]Instituto Madrileño de Estudios Avanzados en Nanociencia (IMDEA Nanociencia), Campus de Cantoblanco, E-28049 Madrid, Spain.
[5]Physical & Materials Chemistry Division, CSIR-National Chemical Laboratory, Dr. Homi Bhabha Road, Pune 411008, India.

Titanium trisulfide (TiS$_3$) has recently attracted the interest of the 2D community as it presents a direct bandgap of ~1.0 eV, shows remarkable photoresponse, and has a predicted carrier mobility up to 10000 cm$^2$V$^{-1}$ s$^{-1}$. However, a study of the vibrational properties of TiS$_3$, relevant to understanding the electron-phonon interaction which can be the main mechanism limiting the charge carrier mobility, is still lacking. In this work, we take the first steps to study the vibrational properties of TiS$_3$ through temperature dependent Raman spectroscopy measurements of TiS$_3$ nanoribbons and nanosheets. Our investigation shows that all the Raman modes linearly soften (red shift) as the temperature increases from 88 K to 570 K, due to the anharmonic vibrations of the lattice which also includes contributions from the lattice thermal expansion. This softening with the temperature of the TiS$_3$ modes is more pronounced than that observed in other 2D semiconductors such as MoS$_2$, MoSe$_2$, WSe$_2$ or black phosphorus (BP). This marked temperature dependence of the Raman could be exploited to determine the temperature of TiS$_3$ nanodevices by using Raman spectroscopy as a non-invasive and local thermal probe. Interestingly, the TiS$_3$ nanosheets show a stronger temperature dependence of the Raman modes than the nanoribbons, which we attribute to a lower interlayer coupling in the nanosheets.

1. Introduction

After the isolation of graphene,[1] a large number of 2D inorganic layered materials have been investigated[2,3] The transition metal dichalcogenide (TMDC) family is amongst the ones that have raised the largest interest in the nanoscience and material science communities as its members present a large variety of electronic properties (ranging from wide band gap semiconductors to superconductors that possess a Peierls instability and display charge density wave transport).[4–9] The semiconducting TMDCs ($MoS_2$, $MoSe_2$, $WS_2$ and $WSe_2$) are especially relevant because of their potential in various types of nanodevices including photodetectors,[10–13] field effect transistors,[14–19] gas sensors,[20] solar cells,[21–25] energy storage devices [26] and field emitters.[27]

Besides the TMDC family, the trichalcogenide family (with a general formula $MX_3$, M = Ti, Zr, Hf, Nb, Ta; X = S, Se, Te) promises the same richness of electronic and optical properties but remains almost unexplored.[28–30] Within the trichalcogenide family members, titanium trisulfide ($TiS_3$) has recently attracted the interest of the 2D community as it has been demonstrated that single layers of $TiS_3$ can be isolated by mechanical exfoliation and field-effect transistors and photodetectors can be realized.[31,32] Unlike Mo- and W-based dichalcogenides (which present a direct bandgap only at the single-layer limit) $TiS_3$ has a direct bandgap of ~1.0 eV which is optimum for photovoltaic and photocatalysis applications.[33–35] Moreover, recent theoretical works predict that the ideal carrier mobility of $TiS_3$ can reach values as high as 10000 $cm^2V^{-1} s^{-1}$.[36] A study of the vibrational properties of trichalcogenides is important to understand their electron-phonon interaction, which plays an important role in the electronic performance of nanodevices and it can even be the main mechanism limiting the charge carrier mobility.[37] But a systematic study of the vibrational properties of $TiS_3$ at different temperatures is still lacking.

In this work, we study the vibrational properties of $TiS_3$ by means of Raman spectroscopy performed at a wide range of temperatures from 88 K to 570 K. Two sets of $TiS_3$ samples, synthesized at different temperatures, were studied: $TiS_3$ nanoribbons (grown at 550 ºC) and $TiS_3$ nanosheets (grown at 400ºC).[32] The samples were characterized by X-ray diffraction, Transmission Electron Microscopy and Electron Diffraction, showing that both sets of samples present high crystallinity. Their Raman spectra showed four prominent peaks corresponding to the excitation of $A_g$-type Raman modes, whose frequencies exhibit a linear temperature dependence over the entire temperature range from 88 K to 570 K. The first order temperature coefficients ($\chi$) associated with each Raman mode has been extracted from the measured temperature dependence. This variation with temperature, similar to that observed in other 2D systems, is mainly attributed to the anharmonic vibrations of the lattice, which includes contributions from the lattice thermal expansion due to anharmonicity. Interestingly, the temperature dependence of the Raman spectrum of $TiS_3$ nanosheets is noticeably stronger than that of $TiS_3$ nanoribbons, which might indicate a lower interlayer coupling for nanosheets than for nanoribbons.

## 2. Results and Discussion

Figure 1(a) shows the $TiS_3$ monolayer structure where the dark grey and yellow spheres refer to the Ti and S, respectively. As pointed out in the introduction, two sets of $TiS_3$ samples were studied in this work: $TiS_3$ nanoribbons (grown at 550 ºC) and $TiS_3$ nanosheets (grown at 400ºC). We address the reader to Ref. [32] and to the Experimental Methods section for more details on the $TiS_3$ synthesis. Figure 1(b) shows an example of a $TiS_3$ nanoribbon (~30 nm thick) and a $TiS_3$ nanosheet (~5 nm thick) deposited onto a $SiO_2$/Si substrate by means of an all-dry transfer technique. Independently of their morphology, $TiS_3$ nanoribbons and nanosheets present similar Raman spectra (see Figure 1(c)) and have the same stoichiometry and crystal structure. . For the temperature dependent Raman spectrum measurements, the as prepared $TiS_3$ nanoribbons and nanosheets powders (mainly containing ribbons and flakes of 100 nm to 300 nm thick)[32] were simply deposited onto SiO$_2$/Si substrates.

Concerning to the space group of MX$_3$ trichalcogenides (M=Ti, Zr, Hf) and (X=S, Se) there are two stable structures: A (ZrSe$_3$-type) and B (TiS$_3$-type). Both structures belong to the P2$_1$/*m* space group (monoclinic).[38] The TiS$_3$ samples have been structurally characterized by using X-ray diffraction (X-RD) in a Panalytical X'pert Pro X-ray diffractometer (CuK$\alpha$ radiation ($\lambda$=1.5406 Å) and all the studied TiS$_3$ samples belongs to the B-variant. Typical XRD patterns of TiS$_3$ nanoribbon samples are shown in Figure 2. A preferential growth in the (001) direction can be easily seen which is common of the MX$_3$ family. Its composition, analyzed by Energy Dispersive X-ray, (EDX Analyser INCA x-sight) at 20keV electron beam energy, results in a stoichiometric ratio, S/Ti =2.9 ± 0.2. We address the reader to Ref. [32] to find the XRD and EDX characterizations of the TiS$_3$ nanosheets samples.

Figure 3(a-b) shows transmission electron microscopy (TEM) images of several TiS$_3$ nanoribbons. Figure 3(c) shows a high resolution (HR) TEM image of a TiS$_3$ nanoribbon showing lines spaced by ~0.41 nm, which matches with the distance along the "*a*" direction of the lattice (JCPDS-ICDD 15-0783) if the sample is observed at an angle of 30º from the (011) crystal plane. Figure 3(d) shows the selected area electron diffraction (SAED) pattern of the TiS$_3$ nanoribbon indicating the high crystalline nature of the sample. Figure 4(a-b) shows typical low magnification TEM images of several TiS$_3$ nanosheets. Figure 4(c) shows a HRTEM image of a TiS$_3$ nanosheet showing lines spaced by ~0.48 nm, which matches very well with the distance along the "*a*" direction of the lattice (JCPDS-ICDD 15-0783) if the sample is observed almost perfectly aligned with the (011) crystal plane. Figure 4(d) shows the SAED pattern of the TiS$_3$ nanosheet indicating the high crystalline nature of the TiS$_3$ nanosheets sample.

The Raman spectrum of TiS$_3$ nanoribbons and nanosheets shows four prominent A$^1_g$ modes which appear at ~172 cm$^{-1}$, 296 cm$^{-1}$, 366 cm$^{-1}$ and 554 cm$^{-1}$ (labelled hereafter as *I*, *II*, *III* and *IV*) due to the in-plane and out-of-plane vibration modes (see Figure 1(b)).[39] Figures 5(a) and (b) show the temperature dependent Raman spectra of a few-layer TiS$_3$ nanoribbon and nanosheet sample recorded at various temperatures from 88 K to 570 K. It is observed that, for both samples, the Raman spectra peak position of all modes shifts towards lower wave number when the temperature is increased from 88 K to 570 K. Similar behavior has been observed for other 2D materials (such as graphene, graphene oxide, MoS$_2$, MoSe$_2$, WSe$_2$ and BP) and it has been attributed to the anharmonic coupling of phonons and the thermal expansion of the lattice that changes the equilibrium positions of the atoms, changing the interatomic forces and thus the phonon energies.[40–44] We also observed that the Full-Width at Half-Maxima (FWHM) for all the Raman modes increased with an increase in temperature. The evolution of the Raman peak position, $\omega$, follows a linear dependence with the temperature:

$$\omega(T) = \omega_0 + \chi T, \qquad (1)$$

where $\omega_0$ is the peak position of the vibration mode at zero Kelvin temperature and $\chi$ is the first-order temperature coefficient of the respective Raman mode. Here, non-linear coefficients have not been considered as they are only relevant at higher temperatures.

Figure 6(a)-(d) show the temperature evolution of the four Raman modes, measured for the nanoribbon (empty symbols) and nanosheet (filled symbols) samples. The first-order temperature coefficients for the different Raman modes $\chi_i$ can be determined from a linear fit. For the TiS$_3$ nanoribbons the first-order temperature coefficients are $\chi_I$ = -0.008 cm$^{-1}$K$^{-1}$, $\chi_{II}$ = -0.018 cm$^{-1}$K$^{-1}$, $\chi_{III}$ = -0.021 cm$^{-1}$K$^{-1}$ and $\chi_{IV}$ = -0.018 cm$^{-1}$K$^{-1}$. For the nanosheets, the first-order temperature coefficients are found to be $\chi_I$ = -0.022 cm$^{-1}$K$^{-1}$, $\chi_{II}$ = -0.025 cm$^{-1}$K$^{-1}$, $\chi_{III}$ = -0.024 cm$^{-1}$K$^{-1}$ and $\chi_{IV}$ = -0.017 cm$^{-1}$K$^{-1}$. The temperature evolution of the Raman modes of TiS$_3$ is more pronounced than that reported for other 2D semiconductors materials (MoS$_2$,[41] MoSe$_2$,[45] WS$_2$,[43] WSe$_2$[45] or black phosphorous [44] which are as summarized in Table 1) and thus our temperature dependent Raman study can be exploited as a non-invasive local probe to measure the temperature in TiS$_3$-based nanodevices.

From Figure 6 it is also clear that the temperature effect on the Raman modes of TiS$_3$ nanosheets is stronger than that observed in nanoribbons. This effect can be attributed to a reduced interaction between the TiS$_3$ layers in the nanosheets sample that would lead to an enhanced in-plane thermal expansion. The mode at ~170 cm$^{-1}$ shows the most pronounced difference between nanosheets and nanoribbons, indicating that it is probably an antiphase out-of-plane mode where the vibration of the lattice substantially changes the interlayer distance during the oscillation. However, to our knowledge, the exact shape of the Raman vibrational modes in TiS$_3$ it is still unclear. In fact the XRD data presented in Ref. [32] show how the interlayer distance for the TiS$_3$ nanosheets is 0.2% larger than that measured for TiS$_3$ nanoribbons. A similar effect of the interlayer interaction on the first-order temperature coefficient has been observed by Calizo *et al*. in graphene samples.[46] They noted how the temperature coefficient varied inversely with the number of graphene layers, being much larger for single layer graphene (no interlayer interaction) than that for highly oriented pyrolytic graphite.

## 3. Conclusions

In summary, we present a study of the vibrational properties of TiS$_3$ by means of Raman spectroscopy over a wide range of temperatures. We studied TiS$_3$ samples with two different morphologies, nanoribbons and nanosheets, which were controlled during their synthesis. Both samples presented high crystallinity as evidenced by their X-ray diffraction, Transmission Electron Microscopy and Electron Diffraction characterization. The Raman spectra of the TiS$_3$ samples showed four prominent peaks corresponding to the excitation of A$_g$-type Raman modes, whose frequencies exhibit a linear temperature dependence over the entire temperature range studied from 88 K to 570 K. This variation with temperature, similar to that observed in other 2D systems, is due to the anharmonic vibrations of the lattice (including contributions from the lattice thermal expansion). Interestingly, the temperature dependence of the Raman spectrum of TiS$_3$ nanosheets is noticeably stronger than that of TiS$_3$ nanoribbons, which we attribute to a lower interlayer coupling for nanosheets than for nanoribbons. Finally, we found that the temperature evolution of the Raman modes of TiS$_3$ is very strong (stronger than that observed in other 2D semiconductors) and thus Raman spectroscopy can be exploited as a non-invasive local probe to measure the temperature in TiS$_3$-based nanodevices.

## 4. Materials and Methods

**Synthesis of TiS$_3$ nanoribbon and nanosheets**. The synthesis of TiS$_3$ nanoribbons and nanosheets has been accomplished by a solid-gas reaction of Ti powder (Goodfellow, 99,5% purity) with sulfur gas provided by the heating of sulfur powder (Merck, 99,99% purity) into a vacuum sealed ampoule at 550ºC and 400ºC, respectively, during 15 days. The procedure is outlined in Ref. [31,32]. TiS$_3$ powders have been structurally characterized by X-ray diffraction in Panalytical X'pert Pro X-ray diffractometer (CuKα radiation (λ=1.5406 Å)). All diffraction peaks can be attributed to the monoclinic phase of TiS$_3$ (P2$_1$/*m* space group).

The as prepared TiS$_3$ nanoribbons and nanosheets samples (composed by ribbons and flakes of 100 nm to 300 nm thick) were simply transferred onto SiO$_2$/Si substrates to characterize their temperature dependent Raman spectra.

**Temperature dependent Raman spectroscopy**. For the temperature dependant Raman spectroscopy, we have used Renishaw inVia confocal Raman microscope with backscattering configuration. All experiments were carried out in Ar ambient at 5mW laser power and 532 nm laser source with identical conditions. For carrying out the temperature dependant Raman spectroscopy experiments, the Raman spectra were recorded by varying theat temperature in the range of 88 K to 570 K using low temperature measurement by Liquid $N_2$ and high temperature cell.

AUTHOR INFORMATION

**Corresponding Authors**

*Andres Castellanos-Gomez: andres.castellanos@imdea.org

* Dattatray J. Late: datta099@gmail.com or dj.late@ncl.res.in

ACKNOWLEDGMENT

Dr. D. J. Late would like to thank Prof. C. N. R. Rao (FRS), JNCASR and ICMS Bangalore (India) for encouragement and support. This work was supported by the Dutch Organization for Fundamental research (NWO/FOM). A.C-G. acknowledges financial support by BBVA Foundation through the fellowship "I Convocatoria de Ayudas Fundacion BBVA a Investigadores, Innovadores y Creadores Culturales" (project: "Semiconductores ultradelgados: hacia la optoelectronica flexible"). Mire group acknowledges to F. Moreno for technical support. The research work was supported by Department of Science and Technology (Government of India) under Ramanujan Fellowship to Dr. D. J. Late (Grant No. SR/S2/RJN-130/2012), NCL-MLP project grant 028626, DST-SERB Fast-track Young scientist project Grant No. SB/FT/CS-116/2013, Broad of Research in Nuclear Sciences (BRNS) Grant No. 34/14/20/2015 (Government of India), partial support by INUP IITB project sponsored by DeitY, MCIT, Government of India and CINT Proposal #U2015A0083 (USA).

**FIGURES**

**Figure 1:**

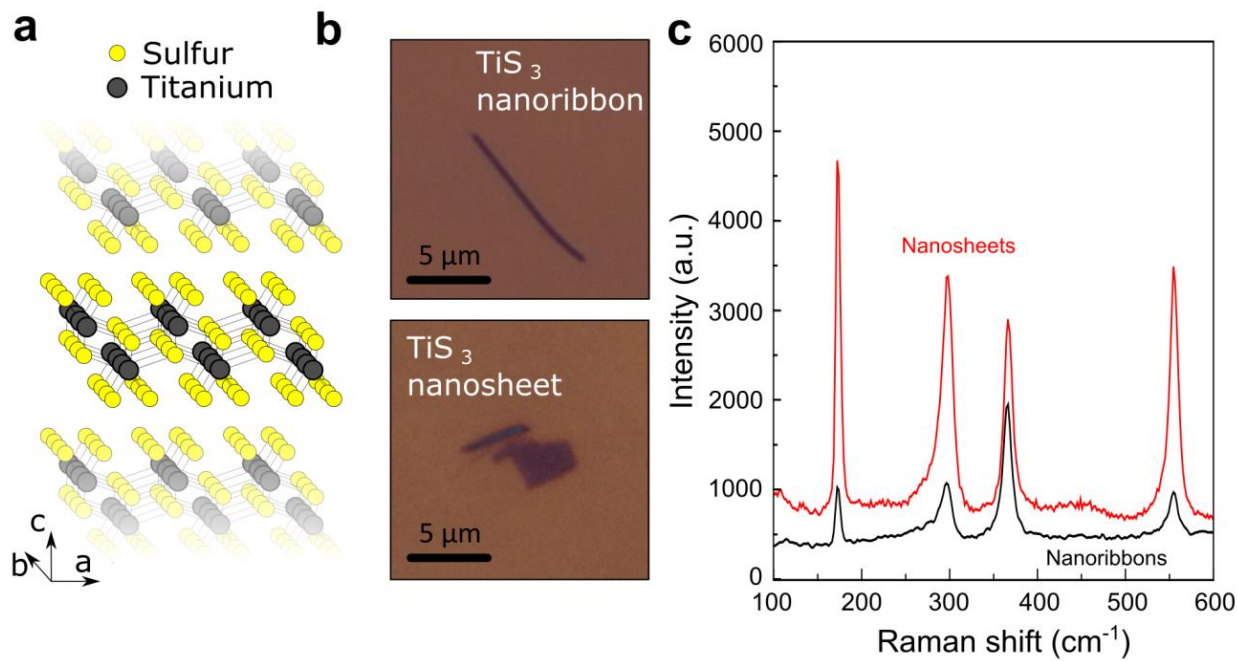

**Figure 1.** (a) Schematic diagram of the TiS$_3$ crystal structure. (b) Optical microscopy images of a TiS$_3$ nanoribbon (grown at 550 ºC) and a TiS$_3$ nanosheet (grown at grown at 400 ºC) deposited onto a SiO$_2$/Si substrate. (c) Raman spectra of TiS$_3$ nanoribbons and nanosheets samples, recorded at room temperature using a 532 nm Ar laser source with 5 mW power.

**Figure 2:**

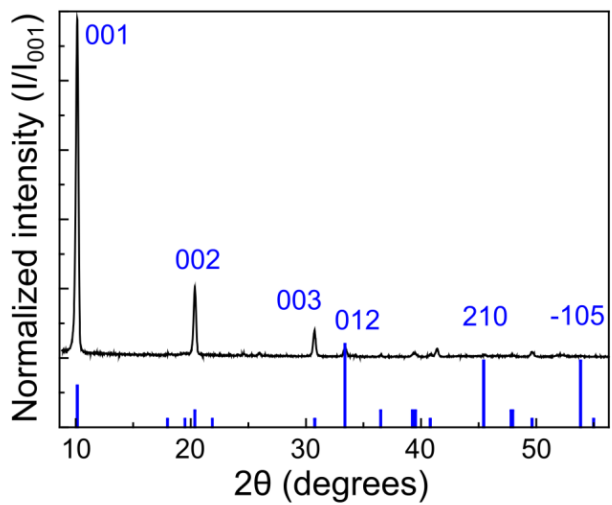

**Figure 2**. XRD diffraction pattern of TiS$_3$ nanoribbons synthesized at 550ºC. Lines corresponding to the monoclinic phase TiS$_3$ (P12/m) as tabulated in JCPDS 015-0783 are included.

**Figure 3:**

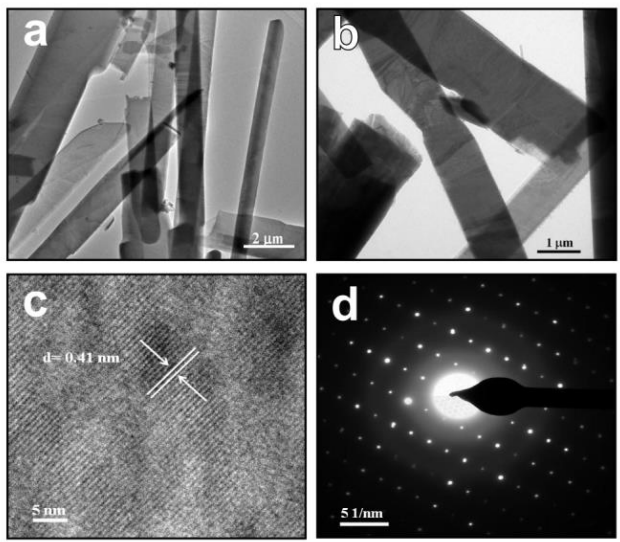

**Figure 3.** (a) and (b) Low resolution TEM images of several TiS$_3$ nanoribbons. (c) HRTEM image of a TiS$_3$ nanoribbon. (d) The selected Area Electron Diffraction (SAED) pattern for the TiS$_3$ nanoribbon.

**Figure 4:**

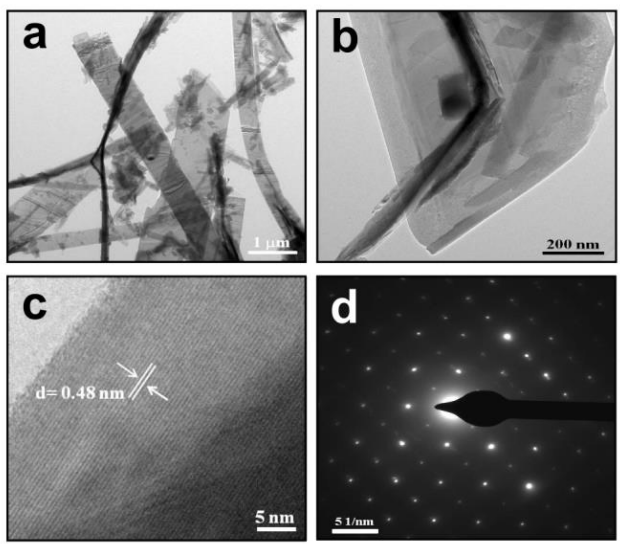

**Figure 4.** (a) and (b) Low resolution TEM images of several TiS$_3$ nanosheets. (c) HRTEM image of a TiS$_3$ nanosheet. (d) A selected Area Electron Diffraction (SAED) pattern for the TiS$_3$ nanosheet.

**Figure 5:**

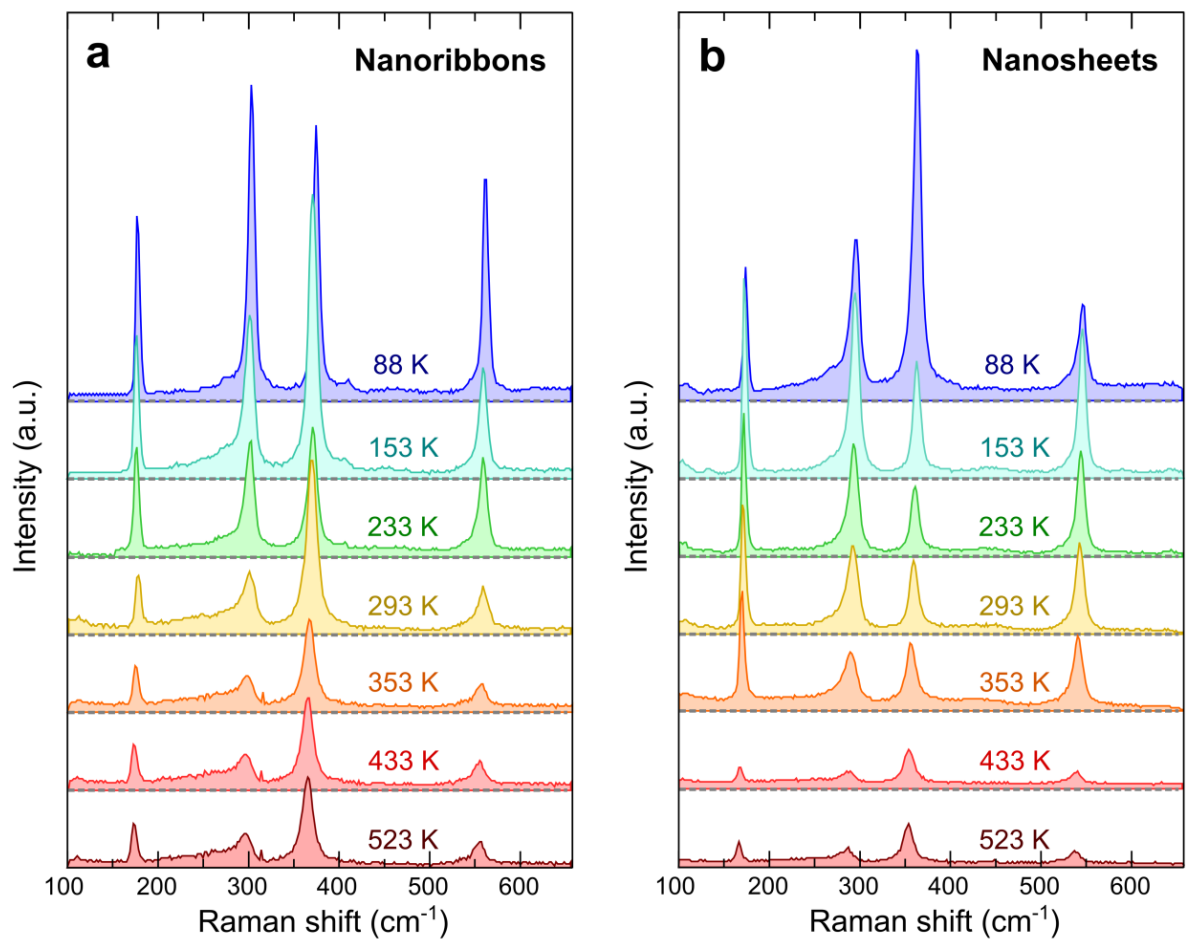

**Figure 5.** Raman spectra of a (a) TiS$_3$ nanoribbon and (b) TiS$_3$ nanosheet sample recorded at various temperatures, acquired with a 532 nm excitation laser.

**Figure 6:**

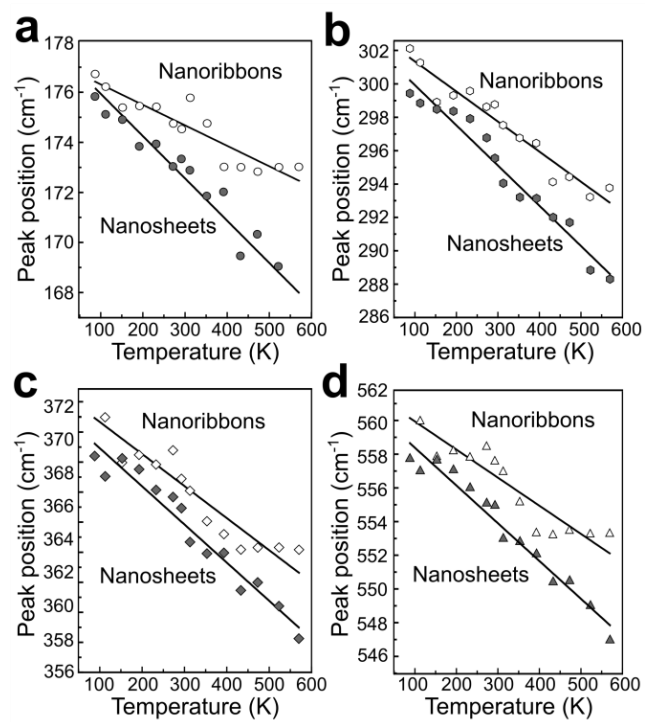

**Figure 6.** Temperature dependence of the four Raman peaks (a-d) measured for TiS$_3$ nanoribbon and nanosheet samples.

**TABLES**.

**Table 1.** Summary of the first-order temperature coefficients reported for the different Raman modes of several two-dimensional materials.

| Material | Temperature Coefficient (cm$^{-1}$ K$^{-1}$) | | | | References |
|---|---|---|---|---|---|
| **Graphene** | $\chi G$ | | $\chi 2D$ | | |
| Single layer Graphene | −0.016 | | −0.026 | | |
| Single layer RGO | −0.029 | | ∼0 | | |
| Six-layer EG | −0.014 | | −0.01 | | [28] |
| Two-layer GNR | −0.030 | | ∼0 | | |
| Six-layer GNR | −0.018 | | −0.026 | | |
| **TMDCs** | $\chi E^1_{2g}$ | $\chi A_{1g}$ | 2LA(M) | A1g(M) + LA(M) | |
| Single layer MoS$_2$ | -0.013 | -0.016 | - | - | [41] |
| Few-layer MoS$_2$ | -0.016 | -0.011 | - | - | [42] |
| Single layer WS$_2$ | -0.006 | -0.006 | −0.008 | −0.01 | [43] |
| Few-layer WS$_2$ | -0.008 | -0.004 | - | - | [43] |
| | $\chi A_{1g}$ | $\chi A^2_{2u}$ | | | |
| Single layer MoSe$_2$ | -0.0054 | -0.0086 | | | [45] |
| Few-layer MoSe$_2$ | -0.0045 | -0.0085 | | | [45] |
| | $\chi E^1 g$ | $\chi A_{1g}$ | A$_{1g}$+LA | 2A$_{1g}$-LA | |
| Single layer WSe$_2$ | -0.0048 | -0.0032 | -0.0067 | -0.0067 | [45] |
| **Black Phosphorous** | $\chi A_{g1}$ | $\chi B_{2g}$ | $\chi A_{g2}$ | | |
| Black Phosphorous | −0.008 | −0.013 | −0.014 | | [44] |
| **Trichalcogenides** | $\chi A_{1g}$ modes | | | | |
| TiS$_3$ nanoribbons | -0.008 | -0.018 | -0.021 | -0.016 | **This work** |
| TiS$_3$ nanosheets | -0.022 | -0.025 | -0.024 | -0.017 | |

19